\documentclass[12pt,preprint]{aastex}

\shorttitle{Lyman-break galaxies}
\shortauthors{Matteucci \& Pipino}

\begin{document}

\title{Lyman-break galaxies: are they young spheroids?}


\author{F. Matteucci\altaffilmark{1} and A. Pipino\altaffilmark{1}}
\affil{Astronomy Department, University of Trieste,
    Via G.B. Tiepolo, 11, I-34127, Trieste, Italy}



\begin{abstract}

We have compared the results from a 
model for the chemical evolution of an 
elliptical galaxy with initial luminous mass of
$2 \cdot 10^{10}M_{\odot}$ and
effective radius of 2 kpc with the recent abundance determinations 
for the Lyman-break galaxy MS 1512-cB58 at a redshift z=2.7276. 
After correcting the iron abundance determination for the presence of 
dust we concluded that  
the observed [Si/Fe], [Mg/Fe], [N/Fe] are consistent with our model
when a galactic age between 20 and 35 Myr is assumed.
Moreover, the [N/O] ratio also suggests the same age.
This age is in very good agreement with other independent studies based 
on the analysis of the spectral energy distribution
suggesting that this object is younger than 35 Myr.
Therefore, we suggest that MS 1512-cB58 is a truly young normal
elliptical galaxy experiencing its main episode of star formation
and galactic wind.
\end{abstract}


\section{Introduction}

Since the discovery of a large population of actively star-forming
galaxies at $3\le z \le 3.5$, identified by their red-shifted Lyman
continuum breaks at 912\AA \,  in the rest-frame (hereafter LBGs,  
see Steidel et
al. 1996a,b), it has became clear that studying their properties as a
whole, as well as their individual spectra, could  allow us to obtain 
information
on star formation histories, ages and metallicities of such high redshift
objects. This information can then be used to compare the LBGs to 
present-day galaxies.  
>From Hubble Deep Field and ground-based far-UV spectra, Steidel et
al. (1996a,b) found that these galaxies are very similar to 
nearby star-forming galaxies. The observation of space density,
star formation rate, masses, physical sizes and morphologies, led
them to the conclusion that the LBGs
they discovered could be the progenitors of the spheroidal components
of present-day luminous galaxies (i.e. elliptical galaxies or spiral
bulges). An argument in favor of ellipticals in clusters as the descendants 
of LBGs is also given by the strong clustering observed for these objects 
(Adelberger et al. 1998; Steidel et al. 1998). On the other hand, 
other authors (Lowenthal et al. 1997; Sawicki and Yee, 1998; 
Sommerville, Primack \& Faber, 2001) favor the idea that LBGs 
can be sub-$L^{*}$ galaxies 
or that they are low mass starbursting objects which later merge to form
more luminous galaxies. A crucial test for the nature of LBGs is to measure 
their masses.
The inferred stellar masses for LBGs are in the range
$M_{*}=10^{9}-10^{11} M_{\odot}$ with an average mass of 
$\sim 10^{10}M_{\odot}$
(Papovich et al.  2001). 
However, these are only lower limits since they measure the 
young starburst component.The inferred ages for the starbursting
populations
range from 30 Myr to 1 Gyr. Again, these are lower limits to the ages of LBGs,
since it cannot be excluded that star formation was
occurring even before the observed 
starburst.

By means of near-IR observations, Shapley et
al. (2001) studied 81, $z\sim 3$ LBGs and found a strong
correlation between the inferred star-formation rates and dust extinction:
younger galaxies have more dust and higher star-formation
rates than older galaxies. Dustier galaxies are also more luminous.
To explain these correlations they proposed
an ``evolutionary sequence'' for LBGs, in which the younger galaxies
evolve into the older, less reddened and quiescent ones.
They derived ages $t_{sf}$ for LBGs ranging 
from several Myr to 1 Gyr,
and best-fit star-formation rates in the range
$\psi (t_{sf})\sim 5-940 \, \rm h^{-2}M_{\odot}yr^{-1}$
with a median $\psi (t_{sf})\sim 45 \, \rm h^{-2}M_{\odot}yr^{-1}$.
The total masses in stars derived by integrating the average star
formation rate over time are  $\ge 10^{10}M_{\odot}$. 
They detected also evidences of winds and outflows driven by SNe and massive
stars, which could enrich the intergalactic medium in metals and energy.

In the following we focus on MS 1512-cB58, the brightest LBG known so far
owing to its gravitationally lensed nature. 
This object 
is at $z=2.7276$ and has a luminous mass of 
$\sim 10^{10}M_{\odot}$, a star formation rate of
$\psi (t_{sf})\sim 40 \,\rm M_{\odot}yr^{-1}$ (Pettini et al. 2001) 
and an effective radius of
$r_{L}\sim 2 \rm \, kpc$ (Seitz et al. 1998), 
for a $\Omega_m= 0.3,\Omega_{\Lambda}=0.7, h=0.70$
cosmology. From the study of MS 1512-cB58  
Pettini et al. (2001) derived also the interstellar medium (ISM) 
chemical abundances.
In particular, they concluded that the abundances of 
O, Mg, Si, P, S are 
$\sim 2/5$ of
their solar values, whereas N, Mn, Fe and Ni are 
underabundant by a factor of $\sim3$. 
Depletion into dust, which is known to be present in MS 1512-cB58,  
probably accounts for some of the Fe-peak element underabundances, 
but this is not
likely to be an important effect for nitrogen.
They indicate that MS 1512-cB58 is young with an age, less
than the typical timescale for the release of N from intermediate mass stars
and Fe from type Ia SNe. They suggest $t_{sf} \le 300 \rm Myr$
while Ellingson et al. (1996) derived an age younger than
$\sim 35$ Myr for the starburst in MS 1512-cB58 by means of 
a fit of the spectral energy distribution (SED). 
Finally, an outflow involving the bulk of the ISM 
and proceeding at a speed of $\sim 255\,\rm Km sec^{-1}$ seems to 
be present in MS 1512-cB58. 
This velocity probably 
exceeds the escape velocity of the galaxy, 
thus suggesting 
that MS 1512-cB58 is probably suffering a galactic wind, although firm 
conclusions on this point are not possible.

Models for the chemical evolution of elliptical galaxies, which are able to 
reproduce the observed properties of local ellipticals, assume that most of 
the 
star formation in these objects took place several billion years ago and 
lasted for a period no longer than 1 Gyr, at the end of which a strong 
galactic 
outflow occurred. These kinds of models are known as ``supernova-driven 
galactic wind models'' for ellipticals and were 
first introduced by Larson (1974).
Supernova-driven wind models including detailed chemical evolution and 
supernova energetics have been later
proposed by Arimoto \& Yoshii (1987) 
and 
Matteucci \& Tornamb\`e
(1987)
and they were aimed at explaining the mass-metallicity relation and the 
color-magnitude diagram of ellipticals.
A multi-zone, more-refined version of these models was then presented by
Martinelli, Matteucci \& Colafrancesco (1988).
These models are also called ``monolithic models'' 
as opposed to 
``hierarchical models'' (see e.g. Kauffmann, 1996), the two main 
competing pictures for galaxy formation.
The main difference between these two scenarios of galaxy formation is 
that in the first case ellipticals formed in a short timescale (1-2 Gyr)
and at very high redshifts (z $\ge$ 3) and 
evolved passively
afterwards, whereas in the second case galaxies assembled in a 
hierarchical fashion.
In particular, the hierarchical process took place over a large time interval 
with the consequence of having 
active star formation until recently
in large ellipticals.

The aim of this letter is to 
test whether the observed chemical properties 
of LBGs, and in particular of MS 1512-cB58, can be explained 
under the assumption that LBGs are normal elliptical galaxies forming at 
high redshift, and
to set independent constraints on the age of this galaxy.
In particular, abundance ratios, especially those between elements formed 
on different 
timescales, can set important constraints on the age and the nature 
of high-redshift objects.
In Section 2 we describe the chemical evolution model we adopt, in 
Section 3 we present the model results and their comparison with the 
observed abundances.
Finally in Section 4 some conclusions are drawn.

\section{The Model}

The elliptical galaxy model we adopted here 
is a multi-zone model
(Martinelli, Matteucci \& Colafrancesco, 1998; Pipino et al. 2002), in which 
the gas collapses into the potential well of a dark matter halo 
at high redshifts ($z \ge 3$) on free-fall timescales. 
The star formation rate is very efficient and  lasts
until the energy injected into 
the ISM by stellar winds and supernovae (Ia,b and II)
triggers a galactic wind which develops first in the most external regions 
and later in the more internal ones. 
As a consequence, while the outer regions 
are suffering an outflow the inner ones are still actively forming stars. 
This kind of model can account for the abundance and color
gradients observed in 
ellipticals (Menanteau, Jimenez \& Matteucci, 2001). 
Generally, after the onset of the galactic wind 
no star formation is assumed to take place in that particular
region, thus supernovae (SNe)
of type II and Ib, 
which originate from massive stars, no longer occur whereas SN Ia 
continue to explode until the present time.
In fact, type Ia SNe are believed to originate from white dwarfs 
in binary systems. Therefore, the evolution of the galaxy, after the 
onset of the wind, is dominated by type Ia SNe which inject 
into the ISM heavy elements
(mostly iron) and thermal energy.

For the
equations of chemical evolution and related details, we direct
the reader to the above mentioned papers,
while here we recall only the main assumptions of the model.
The initial mass function (IMF) is expressed by the classic Salpeter
(1955) law: $\varphi(m) \propto m^{-(1+1.35)}$, defined in the mass
range $0.1-100M_{\odot}$.
Given the uncertainties in the knowledge of the mass of this object we
assumed several values for the initial luminous mass (gas plus stars)
of the galaxy.
In this letter we present three cases, 
with initial luminous masses $10^{9}$, $2 \cdot 10^{10}$ and 
$10^{11}M_{\odot}$
and effective radii of 0.5, 2 and 3 kpc (see Matteucci, 1992), respectively.
Each galaxy is divided in several concentric shell roughly 0.5 kpc wide.

The star formation rate (SFR) $\psi(t)$ is related to the gas
mass by the following equation:
\begin{equation}
\psi(t)=\nu M_{gas}(t)
\end{equation}
with the efficiency of star formation $\nu$
being  in the range 15--12 $Gyr^{-1}$ 
(see Pipino et al. 2002).
In this formulation the  
star
formation rate is maximum at the beginning and then  decreases
with time, as shown in Figure 1 (where the average star formation rate
inside the optical radius of each galaxy is plotted), 
and strongly depends upon 
the galaxy mass. 
The potential energy of the gas is computed by assuming that the galaxy 
is surrounded by a massive and extended 
dark matter halo with $M_{dark}= 10\cdot M_{lum}$ and ${r_L \over r_D}=0.1$,
where $r_{D}$ represents the radius of the dark matter core
(Bertin et al., 1992).
The onset of the galactic wind in each zone depends on the time at which 
the energy, restored by SNe,
exceeds the gas binding energy. At this time, $t_{GW}$, all the gas present 
in the galaxy starts to be lost. Clearly $t_{GW}$ is extremely sensitive 
to the assumptions made about the feedback between SNe and ISM. 
The detailed calculation of
the ISM thermal energy due to stellar winds and SNe
can be found in Pipino et al. (2002).
In that paper we adopted new energetic prescriptions for SNe
taking into account the most recent results concerning the evolution of a 
SN remnant in the ISM. In particular, the most crucial parameter 
in order to compute the amount of energy transferred by a single 
SN into 
the ISM is the cooling time of the remnant. This is the time at which 
radiative losses cannot be
neglected and most of the energy deposited by the shock wave into the ISM, 
during the adiabatic phase, is lost radiatively.
In particular, for type II supernova remnants we adopted the Cioffi, Mc Kee \&
Bertschinger (1988) cooling time, which is a function of density 
and metallicity of
the ISM.  On the other hand, we assumed that SN Ia
can transfer all of their initial blast wave
energy into the ISM. In fact, since type Ia SNe originate
from long living systems, their explosions occur in a 
medium already heated by SNII and therefore the cooling is neglible
(see Recchi et al. 2001).  
Under these assumptions the $2 \cdot 10^{10}M_{\odot}$ model
develops a galactic wind in the most external region 
(roughly 3 kpc from the center) at $\sim 31 \rm Myr$
and subsequently in the inner region (0-1 kpc) at 
$\sim 200
\rm Myr$ after the major episode of star formation has started.
This situation is certainly compatible with what 
observed in MS 1512-cB58, where the outflowing gas is probably located
in front of the star forming region at a distance of a few kpc 
(Pettini et al. 2001).
The amount of stars formed at the time of the onset of the external wind 
is $\sim 1.1\cdot 10^{10}M_{\odot}$ all over the galaxy with the 
stars being more concentrated towards the center.
The models for $10^{9}$ and $10^{11}M_{\odot}$ do not seem to reproduce
the properties of MS 1512-cB58 since they predict too low and 
too high star formation rates at 
any galactic time, respectively, as evident in Figure 1.
They also develop a wind in the most 
external regions too early and too late, respectively.

By means of our model we can predict the evolution of the 
abundances in the gas
of H, He, N, O,
Mg, Si, Fe versus time and metallicity. We adopted different sets of stellar 
yields in different stellar mass ranges: in particular, for massive 
stars ($M >
8M_{\odot}$) the yields of Woosley \& Weaver (1995), for low and 
intermediate mass stars ($0.8 \le M/M_{\odot} \le 8$) the yields
of Renzini \& Voli (1981) and for type Ia SNe the yields of Nomoto
et al. (1984).
It is worth noting that
the model described above has been already compared with the observational 
properties of local ellipticals (see Pipino et al. 2002 and references 
therein),
and
they reproduce both colors and stellar chemical abundances of normal 
ellipticals.

\begin{figure}
\plotone{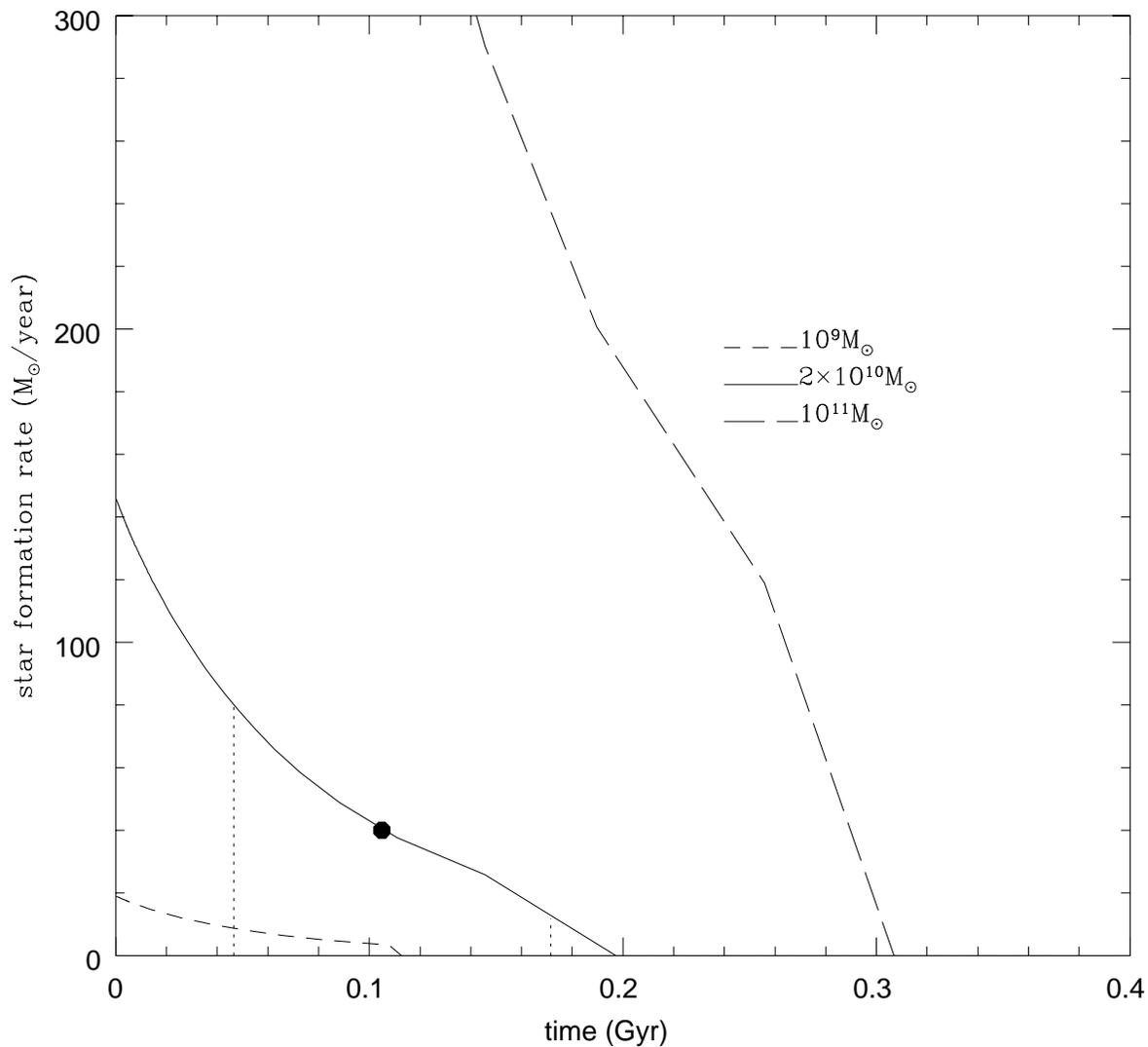}
\caption{The average star formation rates predicted by our models
until the onset of the galactic wind inside the optical radius of each galaxy.
The dotted lines enclose the range 
for the possible values of the star formation rate and the full 
dot represents 
the value adopted 
by Pettini et al. (2001). 
>From the comparison between observed and predicted star formation rate
for the $2 \cdot 10^{10}M_{\odot}$ galaxy
we suggest an age for MS 1512-cB58 of 
$\sim 100 \pm 60 \rm $Myr. \label{fig1}}
\end{figure}

\section{Model results}
In Figure 1 we show the predicted and observed star formation rates for 
MS 1512-cB58. The value quoted by Pettini et al. (2001) is
40$M_{\odot}yr^{-1}$ but we can consider a range of possible values
of 20--80 $M_{\odot}yr^{-1}$, given the uncertainties connected with the 
derivation of the star formation rate. From the comparison between 
the predictions of the $2 \cdot 10^{10} M_{\odot}$ model (the best model) 
and the observed value,
we infer an age for the galaxy inside one $r_L$ of 
$\sim 100 \pm 60$Myr.
In Figures 2 and 3 we show the predicted  [N/Fe], [Mg/Fe] and [Si/Fe]
versus [Fe/H] and time, respectively, and compare them with
the observed ones in MS 1512-cB58. 
The observed values have been corrected for possible dust depletion of Fe,
although the underabundance of Fe-peak elements (relative to the 
undepleted S) cannot be entirely due 
to this effect. In fact, if that would be the case we should expect
the abundances of Mg and Si to be also depleted, although to a lesser extent, 
relative to S. This is not what is observed since the ratios of Mg and Si 
relative to S 
are solar, as discussed by Pettini et al. (2001).  
We assumed that roughly half of the total Fe mass is hidden 
in grains (Pettini, private communication). 
Therefore, the observed Fe abundance has been increased by roughly
a factor of 2. 
As Figure 2 shows, the agreement between model predictions and observations 
is good for the [el/Fe] vs. [Fe/H] relationships, except for Mg which is 
predicted to be lower than observed.
However, this is expected since all the available yields from massive stars 
underestimate the solar Mg abundance (e.g. Thomas, Greggio \& Bender, 1998).
Unlike the star formation rate, the abundance ratios in the gas
are quite insensitive 
to the galaxy mass, whereas they depend strongly upon the nucleosynthesis,
The IMF and timescales for the 
production of the different elements.

In Figure 3, where the abundance ratios are plotted as functions of 
galactic age and redshift, we indicate the observational values 
by dotted lines and see 
where these lines intercept the predicted curves. The interception gives an 
indication of the possible age of MS 1512-cB58. For [Mg/Fe] and 
[Si/Fe] 
the estimated galactic age can be as low as 
$\sim 10$ Myr,
whereas for the [N/Fe] ratio the inferred age 
is $\sim 35$ Myr, in excellent agreement with other independent 
estimates (Ellingson et al. 1996). From the [$\alpha$/Fe] ratios we
cannot exclude ages larger than 10 Myr but we can exclude ages larger than 
300 Myr since the ratios at these times are considerably lower than 
observed. 
The high values of the [$\alpha$/Fe] ratios indicate, in fact, 
that the object 
is younger than the typical timescale for the enrichment from SNe Ia (0.3-0.5
Gyr for ellipticals, Matteucci \& Recchi, 2001). 
Nitrogen is also 
a good cosmic clock since it is mainly produced by low and 
intermediate mass stars, although massive stars contribute as well
for a small amount of this element. 
As a consequence, N starts to be produced in a non 
neglible 
way only after 35 Myr (the lifetime of a 8 $M_{\odot}$ star) while 
its bulk appears only after 300 Myr (the lifetime of a 2.7 $M_{\odot}$).
The ratio of N relative to any $\alpha$-element is even more sensitive
to age then the [$\alpha$/Fe] ratio and this is because  N is mainly
a secondary element and therefore sensitive to the overall metallicity.
The age-[Fe/H] relationship predicted by the model for the 
$2 \cdot 10^{10} M_{\odot}$ galaxy suggests that [Fe/H]=-0.85$\pm 0.1$ dex 
(the dust-corrected 
iron abundance) is reached at an age between 20 and 35 Myr. 
It is worth noting that we obtain a good agreement with observations 
also for the
predicted [N/O] ratio which is independent of dust depletion. In fact, 
we considered the oxygen abundance measured for MS 1512-cB58 by Teplitz 
et al. (2000) 
([O/H]=-0.35 dex) together with the N abundance by Pettini et al. (2001) 
and compared the resulting [N/O] vs. [O/H] and time.
In Figure 4 we 
show the two plots for the model galaxies which show that
the observed [N/O] ratio is compatible with 
an age between 25 and 30 Myr, in good agreement with that estimated from 
the [N/Fe] ratio.

\begin{figure}
\plotone{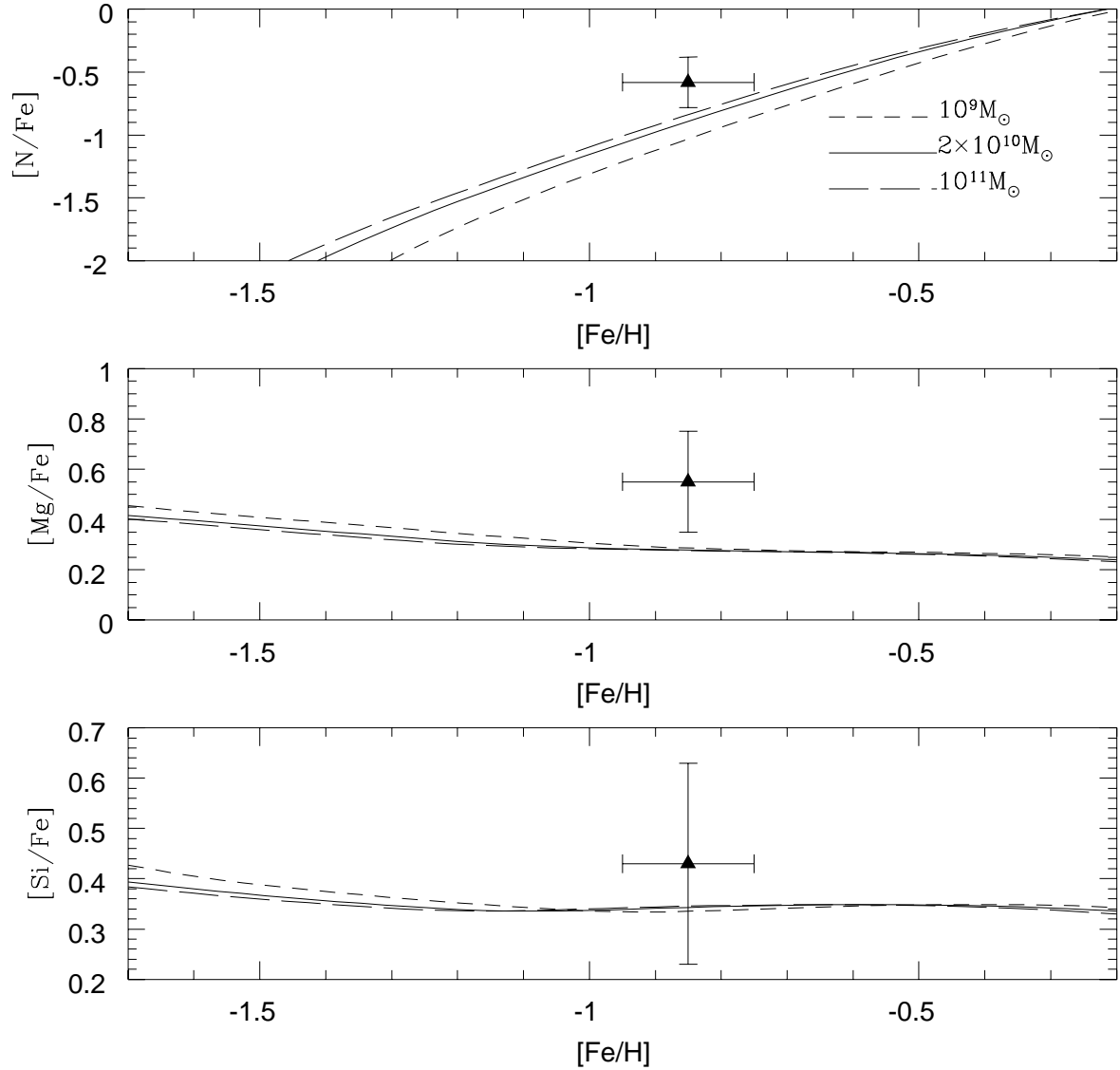}
\caption{Predicted abundance ratios relative to the Sun versus [Fe/H].
The assumed solar abundances are the same adopted in Pettini et al. (2001).
The observed values, dust corrected, are indicated. The luminous initial 
masses are indicated. 
\label{fig2} }
\end{figure}

\begin{figure}
\plotone{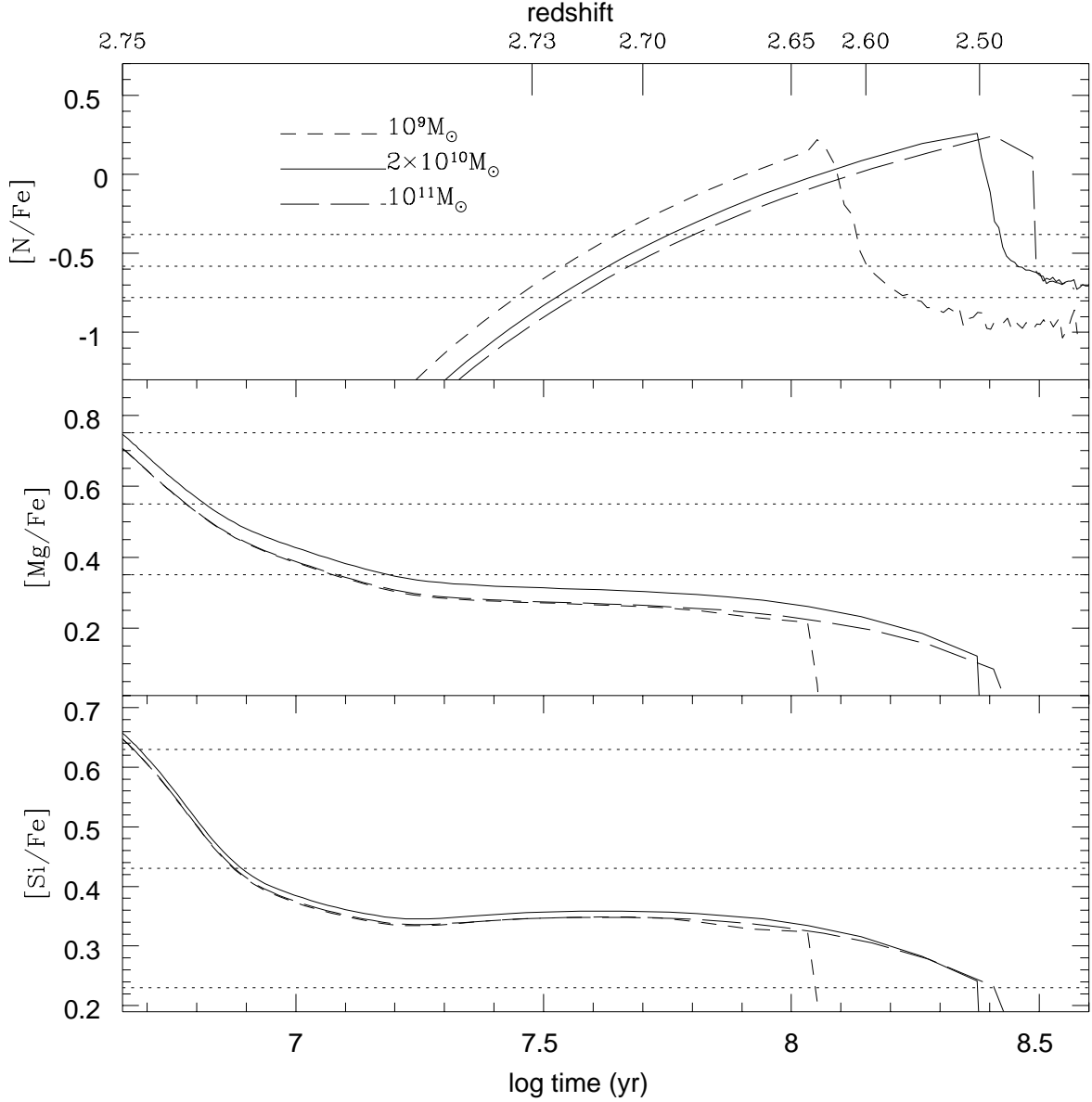}
\caption{Predicted abundance ratios relative to the sun versus galactic age,
for the different model galaxies. 
The models are as in Figure 2. The evolution of the abundance ratios in 
the external region of each model is very similar to that show here 
(inside an optical radius) but it develops a galactic wind earlier than the in
central region (see text).
The observed values and their errors are indicated by the dotted lines.
The redshift is 
also indicated and the
assumed cosmology
is $\Omega_m=0.3$, $\Omega_{\Lambda}=0.7$ and $h=0.70$.
\label{fig3}}
\end{figure}

\begin{figure}
\plotone{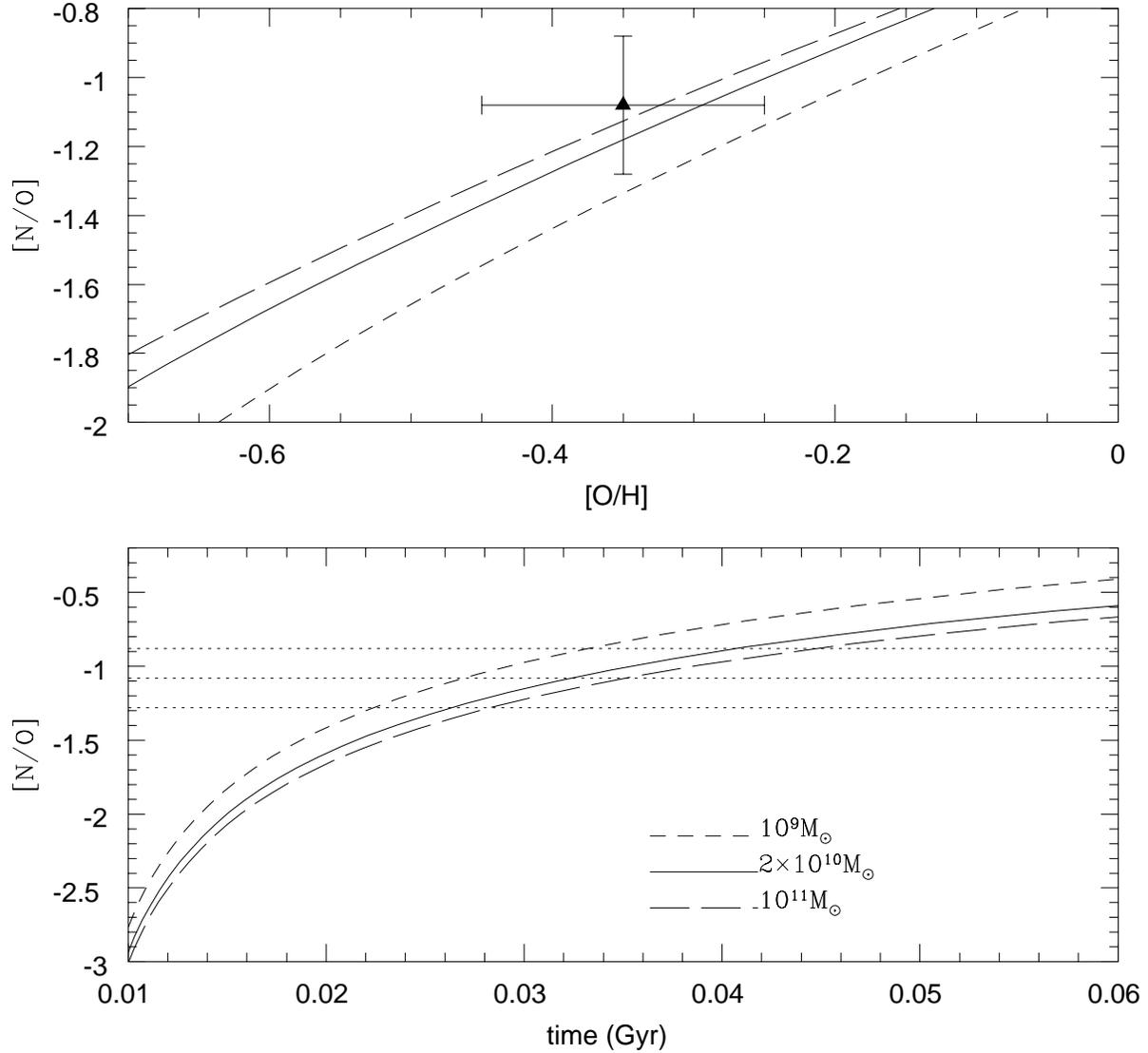}
\caption{Predicted [N/O] ratios versus [O/H] and
galactic age for the same models as in
Fig.2 and Fig. 3.
The observational point is shown as a full triangle.
The N abundance is taken from Pettini et al. (2001) whereas the O abundance
is from Teplitz et al. (2000).
\label{fig4}}
\end{figure}



\section{Discussion and conclusions}
In the previous section we have shown that a SN-driven wind multi-zone
model 
for the 
formation and evolution of an elliptical galaxy with initial luminous mass 
of $2 \cdot 10^{10}M_{\odot}$, which develops a wind in the most 
external region
($\sim 3$ kpc from the center) after $\sim$ 31 Myr, 
can well reproduce the dust-corrected 
abundances measured
by Pettini et al. (2001) in the Lyman-break galaxy MS 1512-cB58.
The fit to the chemical abundances basically 
suggests that MS 1512-cB58 could be a very 
young elliptical galaxy with an age $20 \le t_{sf} \le 35$ Myr, 
experiencing its main starburst and starting to develop an outflow. 
Independent age 
estimates (Ellingson et al., 1996), based on the SED of this galaxy, 
suggest  also an age of 
$\sim$ 35 Myr, in good agreement with our independent estimate.
The abundance ratios represent a strong constraint since they depend mainly 
upon the timescales of the production of the different chemical elements.
On the other hand, the         
star formation rate depends strongly upon the galaxy mass and suggests
a stellar mass of $\sim 1.1\cdot 10^{10}M_{\odot}$ for MS 1512-cB58 
and an age
of $\sim 100 \pm 60$ Myr, but the observed value is affected by the assumed 
Hubble 
constant 
whereas chemical abundances are not, and therefore this age estimate should
be considered less certain than that derived from the abundances.
One of the uncertainties involved in this comparison is that the abundance 
of Fe is very probably affected by dust depletion but we do not know exactly 
how much depletion occurs. We assumed that the observed Fe abundance should 
be increased by a factor of two and this assumption produces 
values for the [$\alpha$/Fe] ratios which are consistent with the [N/Fe] 
ratio. 
On the other hand, if we took
the observed abundances without correction 
the higher [$\alpha$/Fe] ratios would suggest a younger age 
than what we have already inferred whereas
from the higher [N/Fe] ratio a much older one, thus leading to a less 
consistent picture.  
Moreover, our predicted [N/O] vs. [O/H] is also in good agreement with 
the observed values and neither N or O should be substantially
affected by dust depletion.
In any case, dust depletion can only be part of the cause for the
measured Fe 
underabundance, which is mostly due to the young age of the 
galaxy as compared to the typical timescale for Fe production by type 
Ia SNe in ellipticals ($\sim 0.3-0.5$ Gyr, as shown 
by Matteucci \& Recchi, 2001).

In conclusion, this letter supports the idea that LBGs are young 
galaxies which could be the progenitors  
of present day spheroids 
(ellipticals and bulges have very similar properties), although
we cannot predict the precise fate of cB58. We also like to
stress the importance of
using abundance ratios as cosmic clocks for dating high redshift objects.

\section{Acknowledgments} 
We are grateful to Max Pettini for his many important suggestions and to
an anonymous referee for constructive comments.

\end{document}